\documentclass[%
 reprint,
superscriptaddress,
preprintnumbers,
nofootinbib,
 amsmath,amssymb,
 aps,
]{revtex4-1}

\usepackage{graphicx}
\usepackage{dcolumn}

\usepackage{changes}

\def\be{\begin{equation}}
\def\ee{\end{equation}}

\def\l{\left(}
\def\r{\right)}

\renewcommand{\ln}{\mathop{\rm ln}\nolimits}

\newcommand{\Tr}{{\rm Tr}}
\newcommand{\bg}{\begin{gather}}
\newcommand{\eg}{\end{gather}}

\begin{document}

\preprint{INR-TH/2015-031}

\title{Decaying light particles in the SHiP experiment. III.\\ 
      Signal rate estimates for scalar and pseudoscalar sgoldstinos}

\author{K.\,O.\,Astapov}
 \email{astapov@ms2.inr.ac.ru}
\affiliation{Institute for Nuclear Research of the Russian Academy of Sciences,
  Moscow 117312, Russia}%
\affiliation{Physics Department, Moscow State University, Vorobievy Gory,  
Moscow 119991, Russia}

\author{D.\,S.\,Gorbunov}
 \email{gorby@ms2.inr.ac.ru}
\affiliation{Institute for Nuclear Research of the Russian Academy of Sciences,
  Moscow 117312, Russia}%
\affiliation{Moscow Institute of Physics and Technology, 
  Dolgoprudny 141700, Russia}%

\begin{abstract}

For supersymmetric extensions of the Standard Model with light
sgoldstinos---scalar and pseudoscalar superpartners of goldstinos---we
estimate the signal rate anticipated at the recently proposed fixed
target experiment SHiP utilizing a CERN Super Proton Synchrotron beam of 400 GeV protons. 
We also place new limits on the model parameters from a similar analysis of
the published results of the CHARM experiment. 
\end{abstract}

\maketitle

\section{Introduction}
\label{sec:Intro}

Low energy supersymmetry (SUSY) is perhaps the most developed
extension of the Standard Model of 
(SM) of particle physics~\cite{Haber:1984rc,Martin:1997ns}. While, inherent in 
supersymmetry, a technically natural solution to the gauge hierarchy
problem implies the SM superpartners are at or below the TeV energy
scale, other and much lighter new particles can exist as well. In
particular, if supersymmetry is spontaneously broken at not very high
energy scales (see models with gauge mediation of supersymmetry
breaking \cite{Giudice:1998bp,Dubovsky:1999xc} as an example), the
particles from the SUSY breaking sector may show up at quite low
energies. Their effective couplings to the SM particles are
anticipated to be rather weak; therefore a high intensity beam is
required to test the model via production of the new particles.  The
CERN Super Proton Synchrotron (SPS) provides us with a high intensity beam
of 400 GeV protons, and the recently proposed beam-dump Search for Hidden Particles (SHiP) experiment \cite{Anelli:2015pba} (see also
Refs. \cite{Gninenko:2013tk,Bonivento:2013jag}) can perform the task (see
Ref.\,\cite{Alekhin:2015byh} for a comprehensive discussion of the
SHiP physics case).

The purpose of this paper is to estimate the signal rate expected at
the SHiP experiment in supersymmetric models with sufficiently light
particles of {\it the Goldstino supermultiplet}. The latter contains
{\it Goldstino} (the Nambu--Goldstone field, fermion) and its superpartners,
{\it scalar and pseudoscalar sgoldstinos}. While goldstinos are $R$ odd,
sgoldstinos are $R$ even and hence can be singly produced in
scatterings of the SM particles and can subsequently  decay into the
SM particles. Of  particular interest are the sgoldstino decays into
two electrically charged SM particles. These decays 
yield the signature well recognizable
at SHiP \cite{Anelli:2015pba}: 
two charged tracks from a single vertex supplemented with a peak 
in the invariant mass of outgoing particles. 
Sgoldstino couplings to the SM fields are
inversely proportional to the parameter of the order of the
squared scale of SUSY breaking in the whole model. This
unique feature of the Goldstino supermultiplet allows us to probe the
SUSY breaking scale by hunting for
the light sgoldstinos. A preliminary estimate of the sgoldstino signal
for a particular production mechanism and decay channel can be found
in Ref.\,\cite{Alekhin:2015byh}.  
Here we significantly extend that study by considering both scalar
and pseudoscalar sgoldstinos  and by investigating both flavor-conserving and flavor-violating sgoldstino couplings to the SM
fermions, which cover various sgoldstino production mechanisms and
decay modes. 

The paper is organized as follows. Section \,\ref{sec:1} contains
the sgoldstino effective Lagrangian. Section \,\ref{sec:2} and \,\ref{sec:3} are devoted to SHiP phenomenology of scalar and
pseudoscalar sgoldstinos, respectively. Here we discuss direct
(Secs. \ref{subsec:A} and \ref{subsec:Abis}) and 
indirect (Secs. \ref{subsec:B} and \ref{subsec:Bbis}) 
production mechanisms for flavor conserving and flavor
violating sgoldstino coupling patterns. We consider various decay
channels of scalar (Sec. \ref{subsec:C}) and pseudoscalar
(Sec. \ref{subsec:Cbis}) sgoldstinos. We present our results
(Secs. \ref{subsec:D}, \ref{subsec:D}, \ref{subsec:Bbis} and \ref{subsec:Cbis}) as 
estimates of the SHiP sensitivity to the SUSY breaking scale in 
particular supersymmetric variants of the SM. 
In Secs. \,\ref{sec:2} and 
\,\ref{sec:3} we also put new limits on the model parameters by
extending our analysis to the case of the CHARM experiment.  
We conclude in Sec.\,\ref{sec:4} by summarizing the results obtained.

\section{Sgoldstino lagrangian}
\label{sec:1}

If supersymmetry exists, it is spontaneously broken in order to be
phenomenologically viable\,\cite{Haber:1984rc}. The breaking happens
when a hidden sector
dynamics gives a nonzero vacuum expectation value $F$ to 
an auxiliary component of a superfield. The dimension of this parameter is
mass squared, and $\sqrt{F}$ is of order of the SUSY breaking scale. 
The fermion component of this superfield, Goldstino $\tilde
G$, becomes a longitudinal component of the gravitino as a result of
the super-Higgs mechanism making the gravitino massive  (for details see Ref.~\cite{Cremmer:1978iv}). Goldstino 
superpartners, scalar $S$ and pseudoscalar $P$ sgoldstinos gain masses $m_{S,P}$ due to higher order 
corrections from the K\"ahler potential. The masses are largely model dependent and 
 we merely keep them as free parameters in the (sub-)GeV range for this study.

Interaction of the Goldstino supermultiplet with other fields is
suppressed by the parameter $F$~\cite{Cremmer:1978iv}.  
To the leading order in $1/F$, sgoldstino
coupling to SM gauge fields
[photons $F_{\mu\nu}$, gluons $G_{\mu\nu}^a$, where the index $a=1,...8$
runs over $SU(3)$ color group generators] and matter fields (leptons $f_L$, up and
down quarks $f_U$ and $f_D$) at the mass scale above $\Lambda_{QCD}$ but below electroweak symmetry breaking reads \,\cite{Gorbunov:2000th,Gorbunov:2001pd}
\begin{widetext}
\begin{multline}
{\cal L}_{eff}=-\frac{1}{2\sqrt{2}F}\l m_S^2S\bar{\tilde{G}}\tilde{G}+
im_P^2P\bar{\tilde{G}}\gamma_5\tilde{G}\r\label{eef}
-\frac{1}{2\sqrt{2}}\frac{M_{\gamma\gamma}}{F}SF^{\mu\nu}F_{\mu\nu}+
\frac{1}{4\sqrt{2}}\frac{M_{\gamma\gamma}}{F}
P\epsilon^{\mu\nu\rho\sigma}F_{\mu\nu}F_{\rho\sigma}
-\frac{1}{2\sqrt{2}}\frac{M_3}{F}SG^{\mu\nu~a}G_{\mu\nu}^a\\+
\frac{1}{4\sqrt{2}}\frac{M_3}{F}
P\epsilon^{\mu\nu\rho\sigma}G_{\mu\nu}^a G_{\rho\sigma}^a
-\frac{\tilde{m}_{D_{ij}}^{LR~2}}{\sqrt{2}F}S\bar{f}_{D_i}f_{D_j}-
i\frac{\tilde{m}_{D_{ij}}^{LR~2}}{\sqrt{2}F}P\bar{f}_{D_i}\gamma_5f_{D_j} 
-\frac{\tilde{m}_{U_{ij}}^{LR~2}}{\sqrt{2}F}S\bar{f}_{U_i}f_{U_j}-
i\frac{\tilde{m}_{U_{ij}}^{LR~2}}{\sqrt{2}F}P\bar{f}_{U_i}
\gamma_5f_{U_j}\\
-\frac{\tilde{m}_{L_{ij}}^{LR~2}}{\sqrt{2}F}S\bar{f}_{L_i}f_{L_j}-
i\frac{\tilde{m}_{L_{ij}}^{LR~2}}{\sqrt{2}F}P\bar{f}_{L_i}\gamma_5f_{L_j}\;. 
\end{multline}
\end{widetext}
Here $M_3$ is the gluino mass,
$M_{\gamma\gamma}=M_1\sin^2\theta_W+M_2\cos^2\theta_W$ with $M_1$ and
$M_2$ being $U(1)_Y$- and $SU(2)_W$-gaugino masses and $\theta_W$
the weak mixing angle, and  
$\tilde{m}_{U_{ij}}^{LR~2}$ and $\tilde{m}_{D_{ij}}^{LR~2}$ are
left-right up- and down- squark soft mass terms. Lagrangian \eqref{eef}
includes only single-sgoldstino terms; considered in 
Refs.\,\cite{Perazzi:2000id,Perazzi:2000ty,Gorbunov:2000th,Demidov:2011rd}, two-sgoldstino terms are
suppressed by $1/F^2$ and are less promising for testing at the SHiP
experiment. In the interesting range of $F$ here, the gravitino is very
light and can be safely replaced by its Goldstino component $\tilde G$,
entering \eqref{eef}, whenever sgoldstino phenomenology at the
beam-dump experiment is considered.
Hence, sgoldstino couplings to the SM fields are proportional to the
soft supersymmetry breaking parameters of the minimal supersymmetric
extension of the SM (MSSM). 


Sgoldstinos also mix with neutral Higgs bosons as described in
Refs.\,\cite{Dudas:2012fa,Bellazzini:2012mh,Astapov:2014mea}: the 
scalar sgoldstino $S$ mixes with neutral
light $h$ and heavy $H$ Higgs bosons, while pseudoscalar $P$ mixes
with axial Higgs $A$. In what follows we only account for the first
mixing, since the other two do not change the sgoldstino phenomenology
at SHiP for the set of models we investigate. Mixing of the sgoldstino
and lightest MSSM Higgs boson (SM-like Higgs) $h$ can be written 
as\,\cite{Astapov:2014mea}
\begin{equation}
\label{L-mixing}
{\cal L}_{mixing}=\frac{X}{F}\,Sh\,,
\end{equation}
where the mixing parameter $X$ is related to the Higgsino mixing mass
parameter $\mu$, the Higgs vacuum expectation value (vev) $v=174$\,GeV, the
parameter $\tan\beta$ describing the Higgs vev ratio, and  
$SU(2)_W$ and $U(1)_Y$ gauge coupling constants $g_2$ and $g_1$ as
follows  
\begin{equation}
\label{X_mixing}
X = 2\mu^3v\sin{2\beta} +
\frac{1}{2}v^3(g_1^2M_1+g_2^2M_2)\cos^2{2\beta}\,. 
\end{equation}
We consider sgoldstino $S$  to be much lighter than the SM-like
Higgs boson of mass $m_h\approx125$\,GeV.  
Therefore, at low energies the above mixing ensures the Higgs-like 
couplings between the scalar sgoldstino and all the other SM
fields. All the couplings are  
suppressed by the mixing angle 
 \begin{equation}
\label{mix_angle}
\theta =-\frac{X}{Fm_{h}^2}\,.
\end{equation}
 
To illustrate the sensitivity of the SHiP experiment to sgoldstino
couplings, in the next sections we present numerical results for
the set of values of MSSM parameters (the benchmark point in the MSSM
parameter space) shown in Table\,\ref{MSSMpoint}.   
\begin{table}[!htb]
\begin{center}
\begin{tabular}{|l|l|l|l|l|}
\hline
$M_1,$ GeV &$M_2,$ GeV&$M_3,$ GeV&$\mu,$ GeV&$\tan\beta$\\ \hline
100&250&1500&1000&6\\ \hline \hline
$m_A,$ GeV&$A_l,$ GeV&$m_l,$ GeV&$A_Q,$ GeV&$m_Q,$ GeV\\ \hline
1000&2800&1000&2800&1000\\ \hline
\end{tabular}
\caption{\label{MSSMpoint}MSSM benchmark point.}
\end{center}
\end{table}
It is an arbitrary choice, except we suppose that all the model parameters 
take experimentally allowed values and the lightest Higgs boson mass
is 125 GeV. Trilinear soft supersymmetry breaking parameters $A_{l,Q}$
are defined by the relations
$\tilde{m}_{D_{ii}}^{LR~2}\equiv m_{D_i}A_Q$, 
$\tilde{m}_{U_{ii}}^{LR~2}\equiv m_{U_i}A_Q$,
$\tilde{m}_{L_{ii}}^{LR~2}\equiv m_{L_i}A_l$, where we use SM fermion
masses $m_{D_i,U_i,L_i}$; $m_A, m_Q, m_l$ are {\it CP-}odd Higgs boson, squark and slepton masses, correspondingly.


\section{Scalar sgoldstino}
\label{sec:2}

In this section we consider two different production 
mechanisms of the scalar sgoldstino relevant for the SHiP setup. 
The first one is the direct production via hard gluon fusion in proton
scatterings off the target material.  The second one is the production in
decays of mesons emerging due to the proton scattering. 

\subsection{Gluon fusion}
\label{subsec:A}
If the sgoldstino is much heavier than the QCD energy scale of 100\,MeV, it
can be produced directly via gluon fusion. The relevant parts of the
sgoldstino interaction Lagrangians \eqref{eef}, \eqref{L-mixing},
\eqref{mix_angle} read 
\begin{equation}
\label{Sgg}
\begin{split}
{\cal L}_{Sgg} \!=\!\l\!\! \theta\, g^{one-loop}_{hgg}(m_S)-  
\frac{\alpha_s(m_S)\beta(\alpha_s(M_3))}
{\beta(\alpha_s(m_S))\alpha_s(M_3)}
\frac{M_3}{2\sqrt{2}F}\!\!\r 
\\{\times}
SG^{\mu\nu\;a}G_{\mu\nu}^a,
\end{split}
\end{equation}
where the first term, associated with Higgs-sgoldstino mixing
\eqref{L-mixing}, is 
proportional to the Higgs effective coupling to gluons 
(appearing at one-loop level via virtual quark exchanges) \cite{Ellis:1979jy},  
\begin{multline}
g^{one-loop}_{hgg} =
\frac{3}{4}\frac{\alpha_s(m_{S})}{6\sqrt{2}\pi{}v}\Big({\cal
  A}_{1/2}(\tau_t) + {\cal A}_{1/2}(\tau_b) + \\
+ {\cal A}_{1/2}(\tau_c)+ {\cal A}_{1/2}(\tau_s)\Big).
\end{multline}
Here $\tau_{i}=\frac{4m_{i}^2}{m_h^2}$, and loop form factors read
\begin{eqnarray}
{\cal A}_{1/2} & = & 2\tau\left(1+(1-\tau)f(\tau)\right)
\end{eqnarray}
with
\begin{equation}
f(\tau) = \left\{
\begin{array}{ll}
\arcsin^{2}\left({1/\sqrt{\tau}}\right), \;\;\; \tau\ge 1\,,\\
-\frac{1}{4}\log{\frac{1+\sqrt{1-\tau}}{1-\sqrt{1-\tau}}}, \;\;\;
\tau<1\,.
\end{array}
\right.
\end{equation}
The factor in front of the second term in Eq.\,\eqref{Sgg} accounts for the
renormalization group evolution with $\beta(\alpha_s)$ being 
the QCD $\beta$ function. 
Note that both terms in Eq. \eqref{Sgg} are inversely proportional to
supersymmetry breaking parameter $F$. 

\begin{figure}[htb]
    \centering
\includegraphics[width=0.45\textwidth]{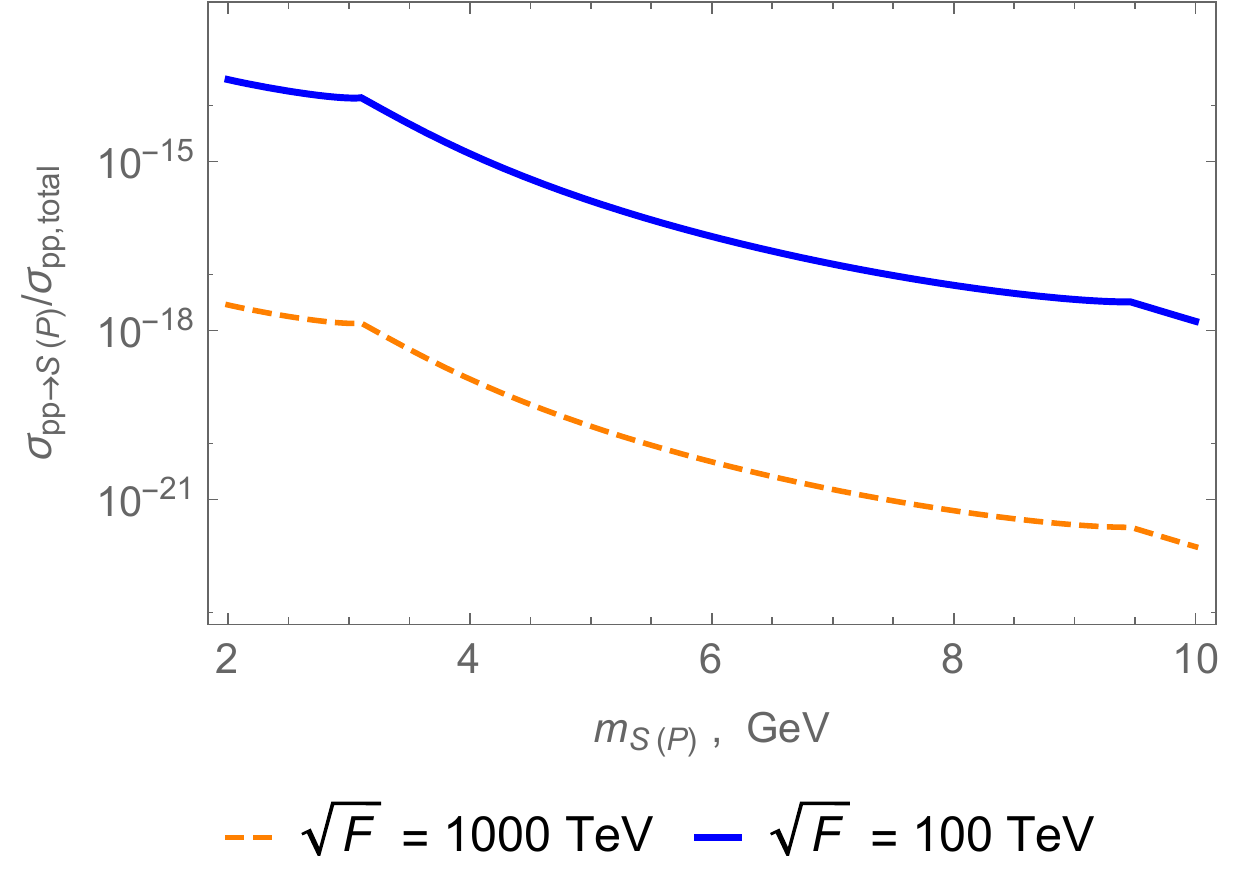}
    \caption{Cross sections of scalar and pseudoscalar sgoldstino
     production in gluon fusion as functions of 
sgoldstino mass. No difference
between scalar and pseudoscalar cases is expected. 
    \label{Sigma}}
\end{figure}

To obtain a reliable estimate of the direct scalar sgoldstino cross
section $\sigma_{pp\to{}S}$, 
we properly rescale the results of Ref.\,\cite{Bezrukov:2009yw}, where 
a coupling similar to Eq. \eqref{Sgg} is responsible for the light inflaton
production at the fixed target experiment with a 400 GeV proton beam. 
For the MSSM parameters from Table\,\ref{MSSMpoint} 
we find the following numerical approximation to the 
cross section as a function of the sgoldstino mass and supersymmetry 
breaking parameter,  
\begin{multline}
\log_{10}(\frac{\sigma_{pp\to{}S}}{\sigma_{pp,\text{total}}})=-15.8666-0.93934\times\left(\frac{m_S}{1\mbox{GeV}}\right)\\
+0.02025\times\left(\frac{m_S}{1\mbox{GeV}}\right)^2+0.00052\times\left(\frac{m_S}{1\mbox{GeV}}\right)^3\\
-4\log_{10}\left(\frac{\sqrt{F}}{100\mbox{TeV}}\right),
\label{fusion}
\end{multline}
where $\sigma_{pp,\text{total}}$ is the total $pp$ cross section for the $400$ GeV
proton beam. This approximation is illustrated in Fig.~\ref{Sigma}
for two reference values of $\sqrt{F}$. Given the
number of protons on target expected at SHiP, about $2\times10^{20}$
\cite{Anelli:2015pba}, one concludes from Fig.\,\ref{Sigma} that the
direct production can provide us with sgoldstinos only in 
the models with a supersymmetry breaking scale below
1000\,TeV, if the MSSM superpartner scale is in the TeV range.

\subsection{$B$ meson decays}
\label{subsec:B}
Scalar sgoldstinos of masses in the GeV range are dominantly produced
by decays of heavy mesons appearing by proton scattering off target. 
In the context of the SHiP experiment the main source of sgoldstinos 
is decays of $B$ mesons.  Decays of charmed mesons are suppressed  as compared to beauty meson decays due to the smallness of the CKM (Cabibbo~--~Kobayashi~--~Maskawa) matrix element in the corresponding amplitude.
The process is described by the triangle diagram with a $t$ quark and $W$ bosons
running in the loop. The sgoldstino is emitted by the virtual
$t$ quark through sgoldstino-top-top coupling \eqref{eef} and 
sgoldstino-Higgs mixing \eqref{L-mixing} (the latter dominates for the
values shown in Table\,\ref{MSSMpoint}). Adopting
the same logic as used in~\cite{Bezrukov:2009yw} for the light inflaton, we
calculate the branching ratio of $B$-meson decay into $S$, 
\begin{multline}
\text{Br}(B\to{}X_{s}S)=\\
=0.3\times\big(\frac{m_t}{m_W}\big)^4\times\left(1-\frac{m_S^2}{m_b^2}\right)^{\!2}
\times(A_Qv+F\theta)^2{\times}\\
\times\left(\frac{100\mbox{ TeV}}{\sqrt{F}}\right)^{\!4}\,,
\label{B-branching}
\end{multline}
where $X_s$ stands for the strange meson 
channel mostly saturated by a sum of pseudoscalar and
vector kaons; $m_b,m_t$ and $m_W$ stand for $b,t$ quarks and $W^{\pm}$ boson masses, correspondingly.
The scalar sgoldstino production cross section is then a product of the
branching ratio above and the beauty cross section evaluated at the
SHiP energy scale as $1.6\times 10^{-7}\times 
\sigma_{pp,\text{total}}$\,\cite{Anelli:2015pba}. 

\begin{figure}[htbp]
    \centering
\includegraphics[width=0.45\textwidth]{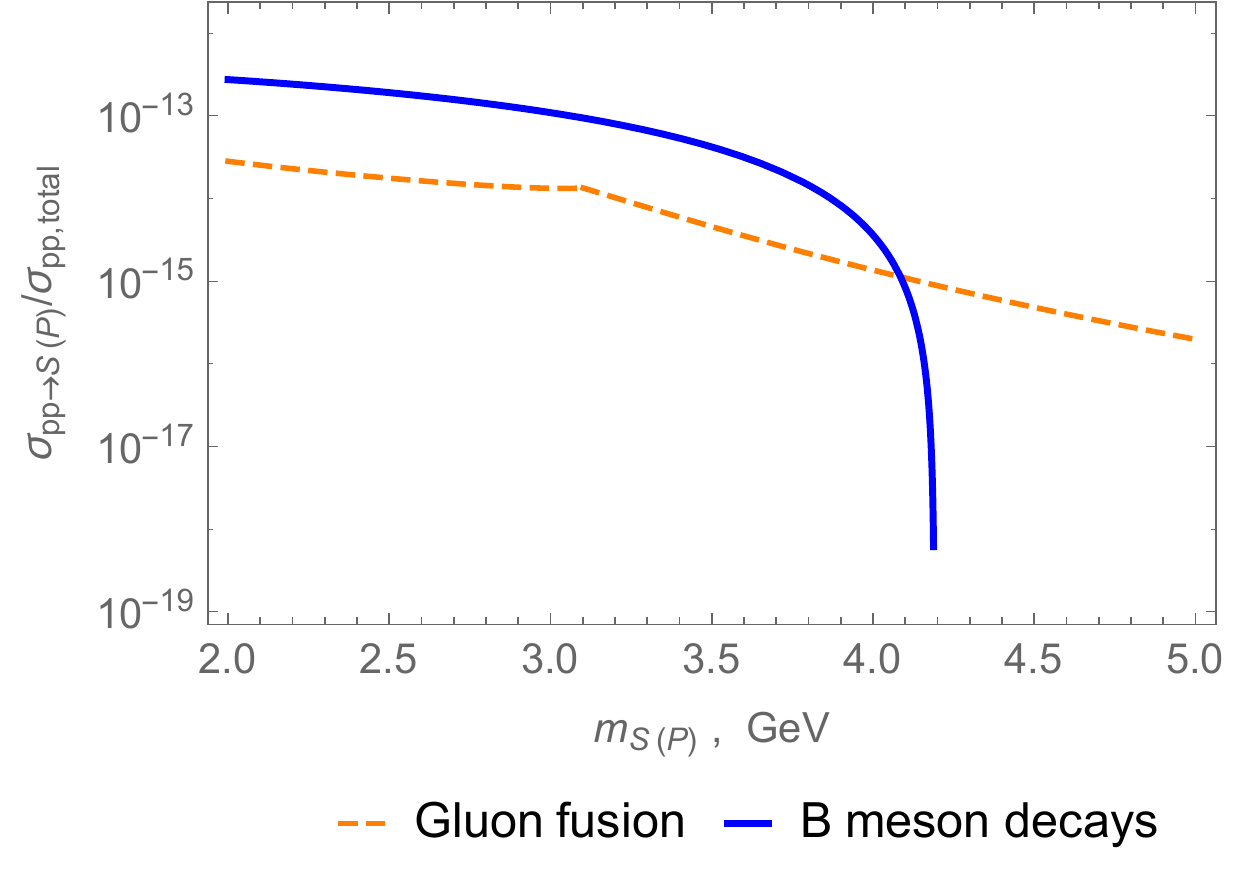}
    \caption{Cross sections of scalar  
sgoldstino production in gluon fusion and in $B$-meson decays for the
model with $\sqrt{F}=100$\,TeV. The same results are valid 
for the pseudoscalar sgoldstino production.
    \label{SigmaBmeson}}
\end{figure}

In Fig.\,\ref{SigmaBmeson} 
we compare the sgoldstino production cross sections provided by the
direct and the indirect mechanisms 
for the same 
value of the supersymmetry breaking parameter, $\sqrt{F}=100$\,TeV. One can
observe from Eqs.\,\eqref{fusion} and \eqref{B-branching} 
that both cross sections scale as $\propto 1/F^2$. Thus, we conclude from
the plot in Fig.\,\ref{SigmaBmeson} that the meson channel  
dominates sgoldstino production when the kinematics allows. In what
follows we concentrate on this case and comment on prospects of
searches for the heavier
sgoldstinos, $M_{S(P)}\gtrsim 4$\,GeV, (available only via the direct
production) in due course.

\subsection{Sgoldstino decay pattern} 
\label{subsec:C}

The sgoldstino is R even and  can decay into 
pairs of SM particles, if it is kinematically allowed. 
For the sgoldstino of the (sub-)GeV mass-range,  the
main decay channels are
$\gamma\gamma$, $e^+e^-,$ $\mu^+\mu^-$, $\pi^0\pi^0$, $\pi^+\pi^-$,
$K^+K^-$, $K^0\bar{K}^0$ (see
Ref.\,\cite{Gorbunov:2000th} for details).  

The rate of sgoldstino decay into photons is described by 
the following expression:
\begin{equation}
\label{Sgammagamma}
\Gamma(S\to\gamma\gamma)=\left(\frac{\alpha(m_S)
\beta(\alpha(M_{\gamma\gamma}))}{\beta(\alpha(m_S))\alpha(M_{\gamma\gamma})}
\right)^2\frac{m_{S(P)}^3M_{\gamma\gamma}^2}{32\pi F^2}.
\end{equation}
Here the dimensionless multiplicative factor accounts for the 
renormalization group evolution of the photonic operator at different
mass scales.
For sgoldstino decays into leptons one finds 
\begin{equation}
\label{Sll}
\Gamma(S\to {}l^+l^-)={m_S^3A_l^2\over 16\pi F^2}{m_{l}^2\over m_S^2}\l
1-{4m_{l}^2\over m_S^2}\r^{3/2}.
\end{equation}

If the scalar is light ($m_S<1.2$\,GeV) then its decay into light
mesons is described by the effective interaction involving the gluonic
operator at a low energy scale, as explained in Ref.\,\cite{Gorbunov:2000th}.  
It yields the rates of sgoldstino decays into mesons, e.g.,   
\begin{multline}
\Gamma(S\to\pi^0\pi^0)={\alpha^2_s(M_3)\over\beta^2(\alpha_s(M_3))}
{\pi m_S\over
4}{m_S^2M_3^2\over F^2}\\ \left( \!\!1\!
-\!{\beta(\alpha_s(M_3))\over\alpha_s(M_3)}
{9\over 4\pi}{B_0\over
m_S}{m_u+m_d\over m_S}{A_Q\over M_3}\!\right)^{\!\!2}\!
\sqrt{1\!-\!{4m_{\pi^0}^2\over m_S^2}},
\label{StoPiPi}
\end{multline} 
where $\beta(\alpha_s)$ is the QCD beta function, $\alpha_s(M_3)$ is the
strong coupling constant evaluated at the scale of $M_3$, 
and the parameter $B_0$ can be expressed via masses 
of kaons and quarks as $B_0=M_K^2/(m_d+m_s)$.

It turns out that for our benchmark point (see Table\,\ref{MSSMpoint})
the contribution to Eq. \eqref{StoPiPi} from the
gluonic operator dominates over the quark operator contribution, so 
we have 
\begin{equation}
\Gamma(S\to\pi^0\pi^0)\approx{\alpha^2_s(M_3)\over\beta^2(\alpha_s(M_3))}
{\pi m_S^3M_3^2\over 4F^2}\sqrt{1-{4m_{\pi^0}^2\over m_S^2}},
\end{equation} 
\begin{equation}
\Gamma(S\to\pi^+\pi^-)=2\Gamma(S\to\pi^0\pi^0)\,,
\end{equation}
and analogously for kaons:
\begin{equation}
\Gamma(S\to{}K^0\bar{K}^0)\approx{4\alpha^2_s(M_3)\over\beta^2(\alpha_s(M_3))}
{\pi m_S^3M_3^2\over 4F^2}\sqrt{\!1\!-\!{4m_K^2\over m_S^2}},
\end{equation}
\begin{equation}
\label{SK+K-}
\Gamma(S\to{}K^+K^-)=\Gamma(S\to{}K^0\bar{K}^0)\,.
\end{equation}
 
For the sgoldstino masses well above the QCD mass scale ($m_S\gg 1$\,GeV) 
the decay into hadrons can be described as 
decay into a gluon pair, which then hadronizes. Its rate reads 
\begin{equation}
\Gamma(S\to{}gg)=
\left(\frac{\alpha_s(m_S)\beta(\alpha_s(M_3))}
{\beta(\alpha_s(m_S))\alpha_s(M_3)}\right)^2
\frac{m_S^3M_3^2}{4\pi{}F^2}\,.
\end{equation}
The multiplicative factor is added above to correct for the renormalization
group evolution of the gluonic operator with mass scale.

Between these mass
ranges, where $m_s\approx1.2 - 4$ GeV, neither description method is reliable where some heavy mesons, as well as multimeson final states, enter the game. However, comparing contributions for relevant hadronic modes calculated within these two approaches,  we observe that deviations do not exceed an order of magnitude. Thus, we conclude
that order-of-magnitude estimates of the sgoldstino lifetime  for the mass interval $1.2-4$ GeV can be obtained by some extrapolation of the chiral theory approach.

\begin{figure}[htb!]
    \centering
\includegraphics[width=0.5\textwidth]{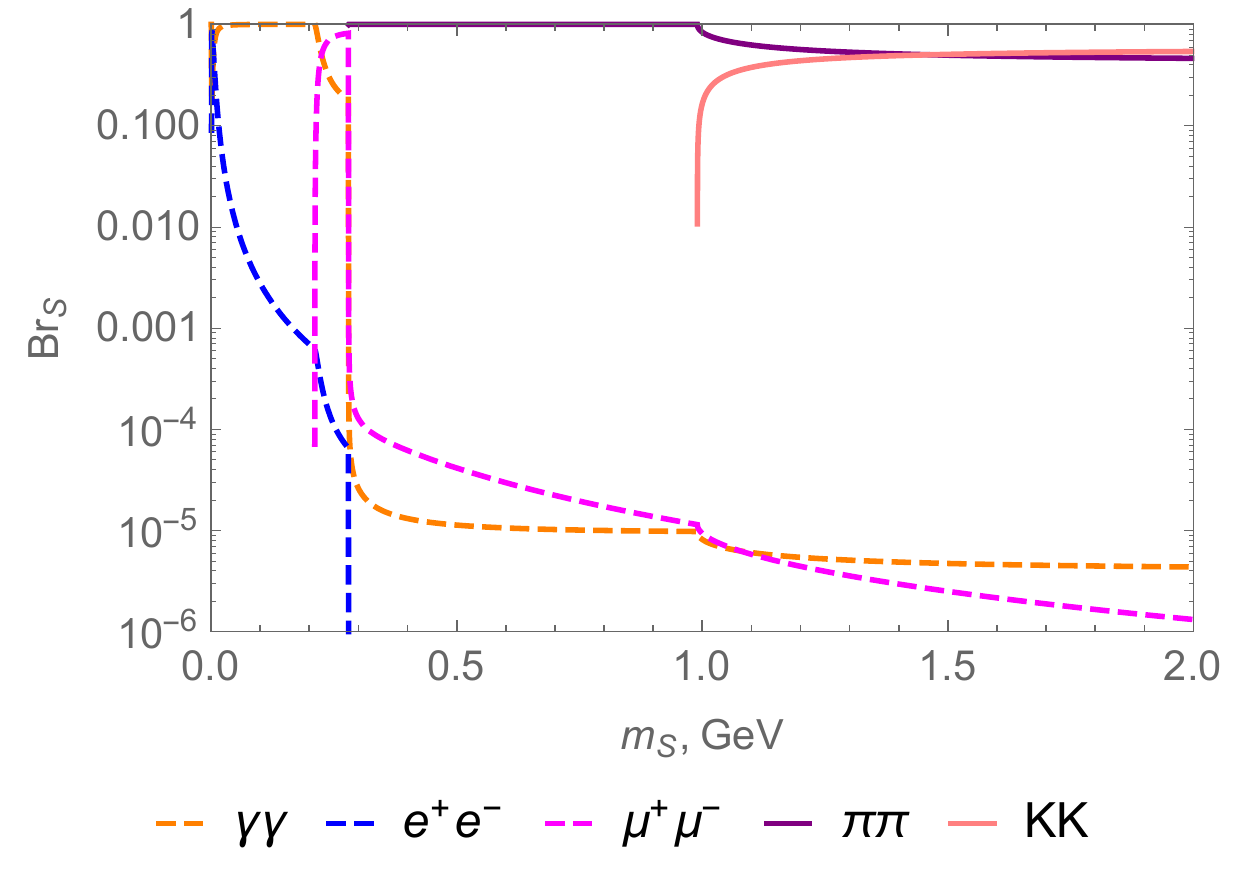}
    \caption{Branching ratios of a scalar sgoldstino.
    \label{SBranching}}
\end{figure}

The branching ratios of the scalar sgoldstino decays 
 for the values of MSSM soft parameters chosen in Table\,\ref{MSSMpoint}
 are shown in Fig.\,\ref{SBranching}. 
Hadronic channels, $\pi\pi$, $KK$, naturally dominate (if
kinematically allowed), while
$\gamma\gamma$ and $\mu^+\mu^-$ give small but noticeable
contributions. 
Decay rates \eqref{Sgammagamma}--\eqref{SK+K-} ensure that each
branching ratio scales as the square of the corresponding soft supersymmetry
breaking parameter, e.g., Br$(S\to\mu^+\mu^-)\propto A_l^2$. 

\begin{figure}[htb!]
   \centering
\includegraphics[width=0.5\textwidth]{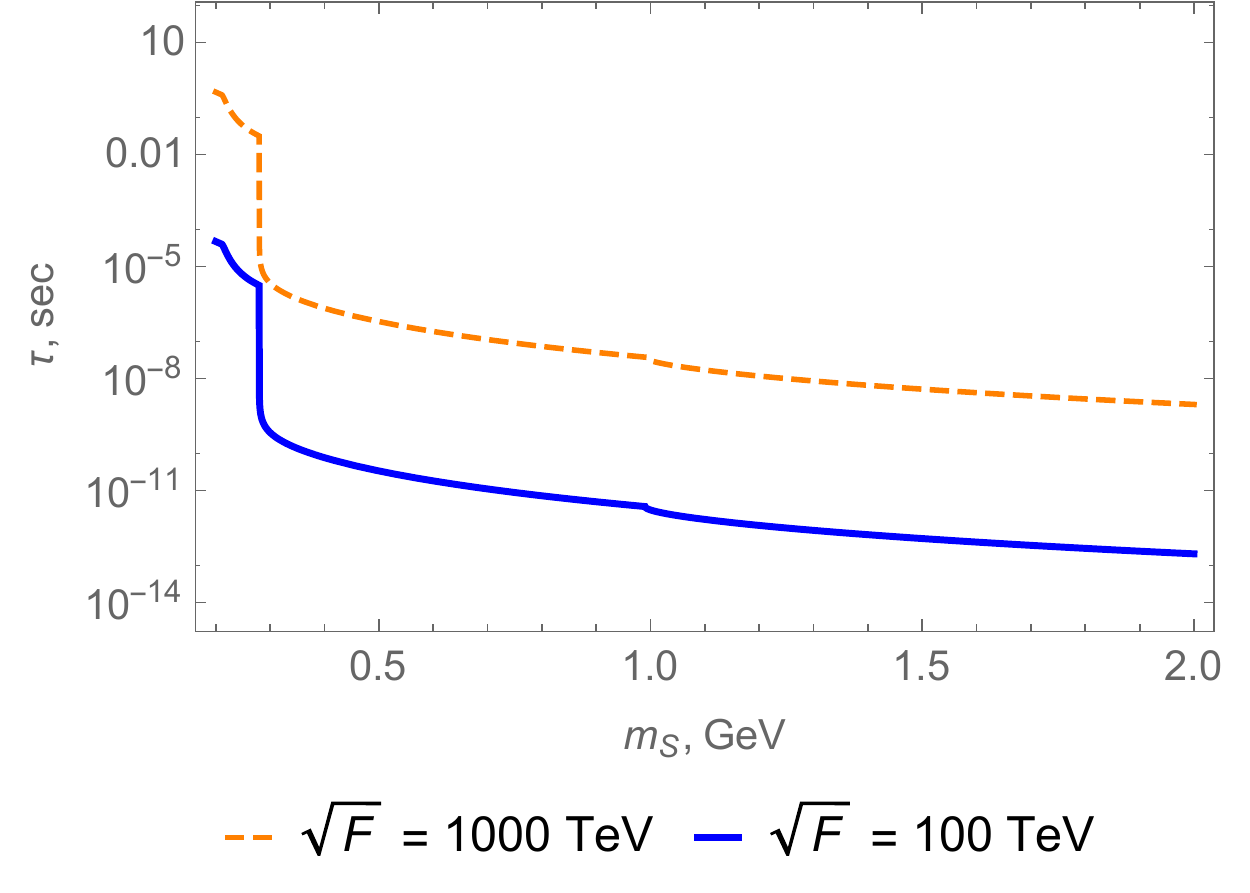}
\caption{Lifetime of a scalar sgoldstino as a function of its mass.
\label{SLifetime}}  
\end{figure}

Heavier sgoldstinos decay mostly into gluons. 
The invisible decay mode $S\to \tilde G \tilde G$ is always
negligible because the corresponding coupling in Eq. \eqref{eef} is
proportional to the sgoldstino mass squared rather than the MSSM soft
terms.  
The sgoldstino lifetime for a set of values of $F$ is 
presented in Fig.\,\ref{SLifetime}. 
To reach the main detector of the SHiP experiment, the sgoldstino has to
cover a distance of about 100 meters\,\cite{Anelli:2015pba}. The
results in Fig.\,\ref{SLifetime} 
suggest that SHiP can be sensitive mostly to the
models with supersymmetry breaking scale of about 100\,TeV and higher.  
The lifetime scales as $\tau\propto F^2$ and as $\tau\propto 1/M_3^2$
since the hadron channel dominates.

\subsection{Sgoldstino signal event rate at SHiP}
\label{subsec:D}
Now we collect all the ingredients required to achieve the main goal
of this article, the estimate of the number of sgoldstino decay
events inside the fiducial volume of the SHiP experiment.  The SHiP
construction is outlined in Ref.\,\cite{Anelli:2015pba}.  The 400 GeV
proton beam fueled by the SPS hits the target and produces bunches of
mesons, which can decay into new particles (sgoldstinos in the case at
hand). The latter can also appear directly from the 
proton-proton collisions (through gluon fusion).  
The detector is placed at a distance
of $l_{sh}=63.8$\,m from the target. The vacuum vessel length is about 
$l_{det}=60$\,m. It
forms a cylinder along the beam axis with an elliptical base of
5\,m$\times$10\,m. The trajectories of electrically charged particles
emerging from the new particle decays can be traced in the detector
volume and their energies and types can be determined by the
registration system utilizing devices arranged at the far end of the
detector.

The differential production cross
section of sgoldstinos, originated from the 
$B$ meson decays, has the following form, 
\begin{equation}
\label{diff}
\frac{d^3\sigma_{pp\to S(P)}}{dpd\theta_pd\phi_p} = 
\int{}d^3\vec{k}\,f(\vec{p},\vec{k})\frac{d^3\sigma_B}{dkd\theta_kd\phi_k} \,,
\end{equation}
where $f(\vec{p},\vec{k})$ is the sgoldstino 
momentum distribution 
normalized to the branching ratio \eqref{B-branching} and
$\frac{d^3\sigma_B}{dkd\theta_kd\phi_k}$ is the
differential production cross section of $B$-mesons in proton-proton
collisions, which is evaluated along the lines of 
Ref.\,\cite{Gorbunov:2015mba}.  In the integral \eqref{diff} the total
values of the 3-momenta and escaping angle of the outgoing particles are
specifically constrained to ensure that the sgoldstino trajectory crosses the
rear end of the SHiP detector, which defines the fiducial volume.  

With the above approximations we estimate the number of signal events
as 
\begin{equation}
N_{\text{signal}}=\frac{N_{\text{POT}}}{\sigma_{pp,\text{total}}} 
\int{}w_{det}\frac{d\sigma_{pp\to{}S(P)}}{dpd\theta_pd\phi_p}d^3\vec{p}\,,
\end{equation}
where the expected number of protons on the target 
is $N_{\text{POT}}=2\times10^{20}$\,\cite{Alekhin:2015byh} 
and $w_{det}$ denotes the probability for the sgoldstino to decay
inside the fiducial volume of the detector,  
\begin{multline}
w_{det}(E_{S(P)}, m_{S(P)}, \sqrt{F})=\exp(-l_{sh}/\gamma{}c\tau_{S(P)})\times\\
\times\left[1-\exp(-l_{det}/\gamma{}(E_{S(P)})c\tau_{S(P)})\right],
\end{multline}
with the sgoldstino gamma factor $\gamma{}(E_{S(P)})=E_{S(P)}/m_{S(P)}$.

\begin{figure}[!htb]
    \centering
\includegraphics[width=0.45\textwidth]{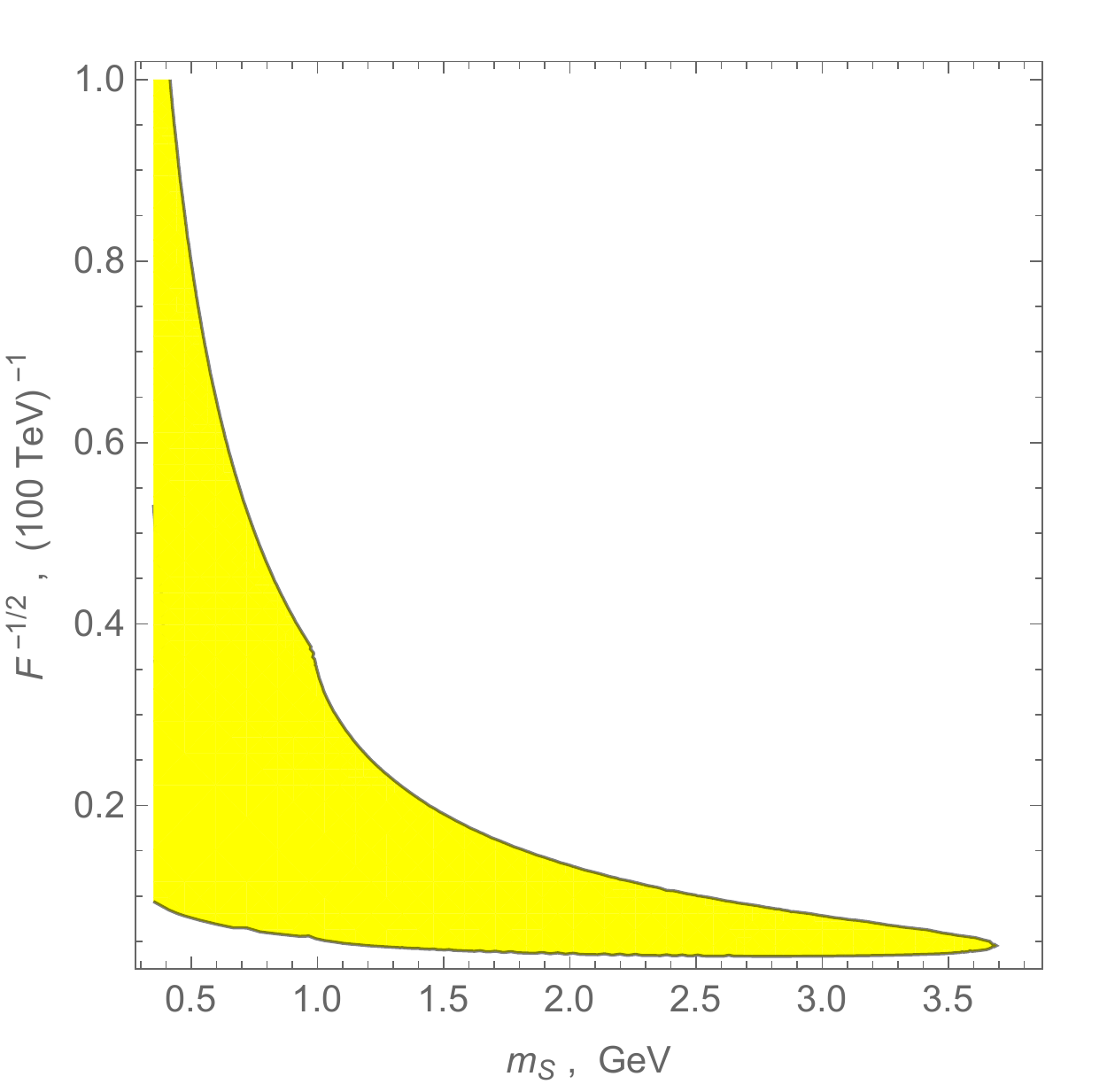}
    \caption{The shaded region  
will be probed at the SHiP experiment. 
    \label{Scalar}}
\end{figure}

In Fig.\,\ref{Scalar} 
we indicate the region in the model parameter space
$(m_{S(P)},1/\sqrt{F})$, where the number of sgoldstino decays inside
the SHiP fiducial volume exceeds 3, $N_{\text{signal}}>3$. That is, if
no events were observed (the background for the two-body decays into charged SM
particles is zero\,\cite{Anelli:2015pba}) the region is excluded at
the confidence level of 95\%, in accordance with the Poisson statistics. 
The upper boundary in Fig.\,\ref{Scalar} is the region where the sgoldstino 
coupling constants $\propto 1/F$ are large enough to initiate 
very fast decay of the sgoldstino before it reaches the detector. 
The lower boundary in Fig.\,\ref{Scalar} 
is the region where the couplings are so small that sgoldstinos
escape from the detector without decay. The number of signal events
here scales with the model parameters as $N_{\text{signal}}\propto
M_3^2\mu^6/F^4$. The region in Fig.\,\ref{Scalar} of the heaviest
sgoldstino reachable at SHiP, $m_S\approx 3.6$\,GeV, is the meeting
point of the lower and the upper boundaries. Here, the sgoldstino decay
length is about 100\,m, which is the scale of both the SHiP detector
length and the distance from the target, $\gamma c\tau\sim l_{det}\sim
l_{sh}$. In this case the number of signal events scales as 
$N_{\text{signal}}\propto \mu^6/F^2$. The scalings of the signal
events imply that models with a higher (as compared to that presented in
Fig.\,\ref{Scalar}) scale of supersymmetry breaking  can be tested if
MSSM parameters $\mu$, $M_3$ are appropriately bigger (as compared to
those presented in Table\,\ref{MSSMpoint}). 

Sgoldstinos of masses $3.6$--$4.2$\,GeV, which can be produced through
$B$-meson decays (see Figs.\,\ref{SigmaBmeson} and 
\ref{Scalar}), seem to be beyond the
SHiP's grip for our choice of MSSM parameters presented in 
Table\,\ref{MSSMpoint}. However, the  signal scaling
with model parameters explained above suggests that sgoldstinos of masses 
above $3.6$\,GeV can be tested 
at SHiP in models with a higher scale of SM superpartners. Finally, from
the results presented in Figs.\,\ref{SigmaBmeson} and \ref{SLifetime} one
can conclude that sgoldstinos of masses above 4\,GeV, which can appear
only via direct production, cannot be tested at SHiP. Both sgoldstino
production and decay are governed by the same ratio $M_3^2/F^2$, and the
sgoldstino lifetime $\tau\propto1/m_S^3$ is too short for the
reasonably high sgoldstino production. One needs much higher
intensity of the proton beam at SPS to probe this region of model
parameter space.

\subsection{Flavor violating}
\label{subsec:E}

In supersymmetric models with nondiagonal sfermion left-right mass terms, 
$\tilde{m}_{D_{ij}}^{LR~2}\propto \!\!\!\!\!\!\!/ \;\,\,\delta_{ij}$, etc,
sgoldstino couplings \eqref{eef} violate flavor symmetry. These flavor-violating terms are additional sources of sgoldstino
production in a beam-target experiment. To illustrate the point, here 
we consider flavor-violating 
decays of (produced by protons on target) $B$ and $D_s$ mesons 
into kaons and light scalar sgoldstinos. 
These processes are
governed by the soft parameters $\tilde{m}_{D_{23}}^{LR~2}$ and 
$\tilde{m}_{U_{12}}^{LR~2}$.  Observation of oscillations in the
$B^0-\overline{B^0}$ and $D^0-\overline{D^0}$ systems and searches for
very rare (within the SM) decays like $B\to\mu^+\mu^-$ 
constrain\, \cite{Ciuchini:2007cw,Arana-Catania:2013pia} possible
flavor violation in the squark sector, which
for our reference point given in Table\,\ref{MSSMpoint} imposes the
following upper limits on the off-diagonal entries, 
\begin{equation}
\label{phen-limits}
\tilde{m}_{D_{23}}^{LR~2}<0.02\,\text{TeV}^2,\;\;\;
\tilde{m}_{U_{12}}^{LR~2}<0.016\,\text{TeV}^2.
\end{equation}

The above flavor-violating interaction terms yield the decay rates
\begin{multline}
\Gamma(B\to{}KS)=\\
=F^2_{B\to{}K}\left(\frac{m_{B}^2-m_K^2}{m_b+m_u}\right)^2\frac{\lambda_{B\to{}KS}}{16\pi^2m_B^3}\frac{\tilde{m}_{D_{23}}^{LR~4}}{F^2}
\end{multline}
and
\begin{multline}
\Gamma(D_s\to{}KS)=\\
=F^2_{D_s\to{}K}\left(\frac{m_{D_s}^2-m_K^2}{m_c+m_u}\right)^2
\frac{\lambda_{D_s\to{}KS}}{16\pi^2m_{D_s}^3}\frac{\tilde{m}_{U_{12}}^{LR~4}}{F^2},
\end{multline}
where 
\[
\lambda_{B(D_s)\to{}KS}=\sqrt{(m_{B(D_s)}^2-m_K^2-m_S^2)^2-4m_{B(D_s)}^2m_K^2},
\]
and dimensionless form factors $F_{B\to{}K}$  and $F_{D_s\to{}K}$  
are given in Ref.~\cite{Palmer:2013yia}. 

\begin{figure}[!htb]
\centering
\includegraphics[width=0.45\textwidth]{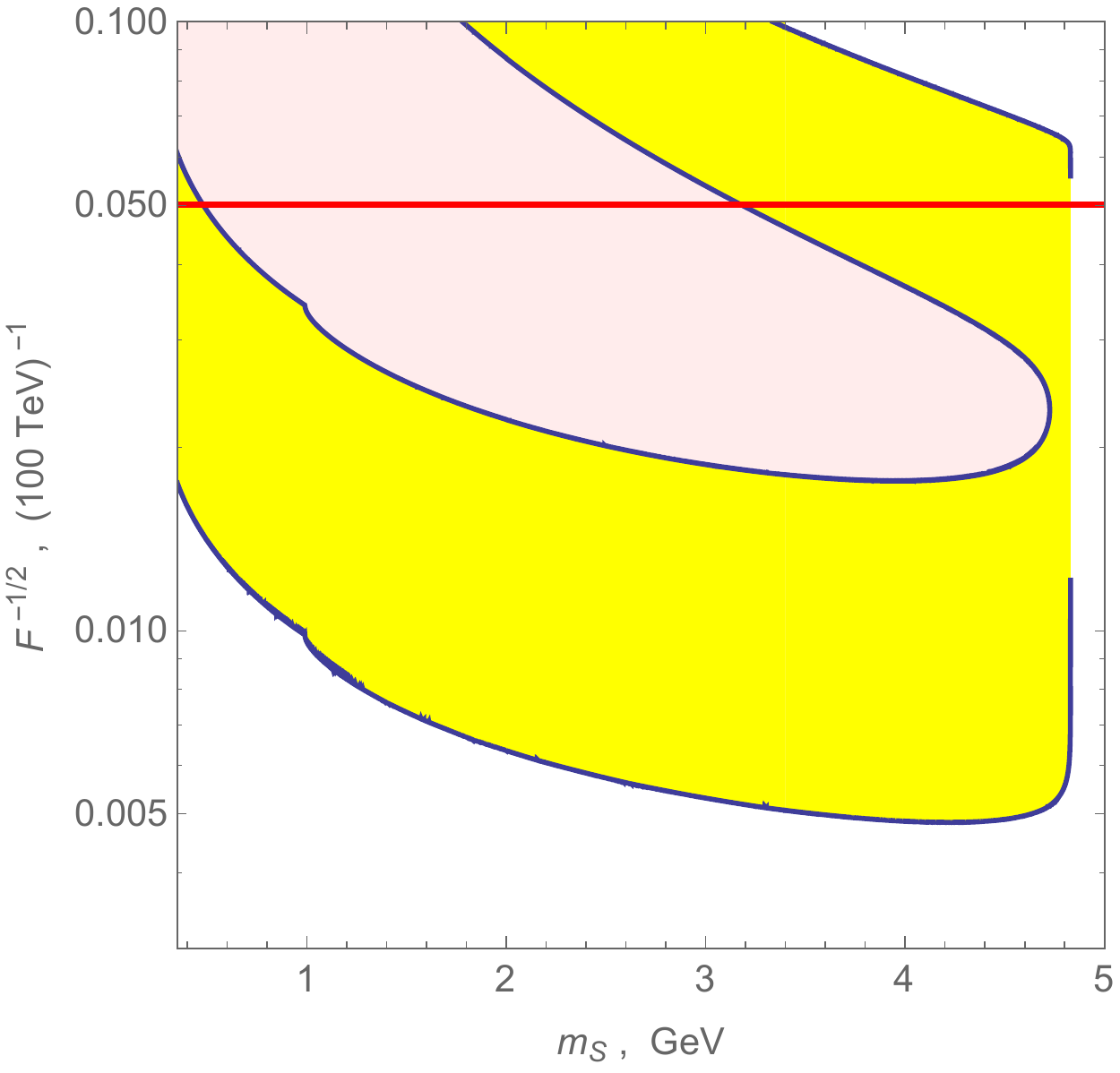}
\includegraphics[width=0.45\textwidth]{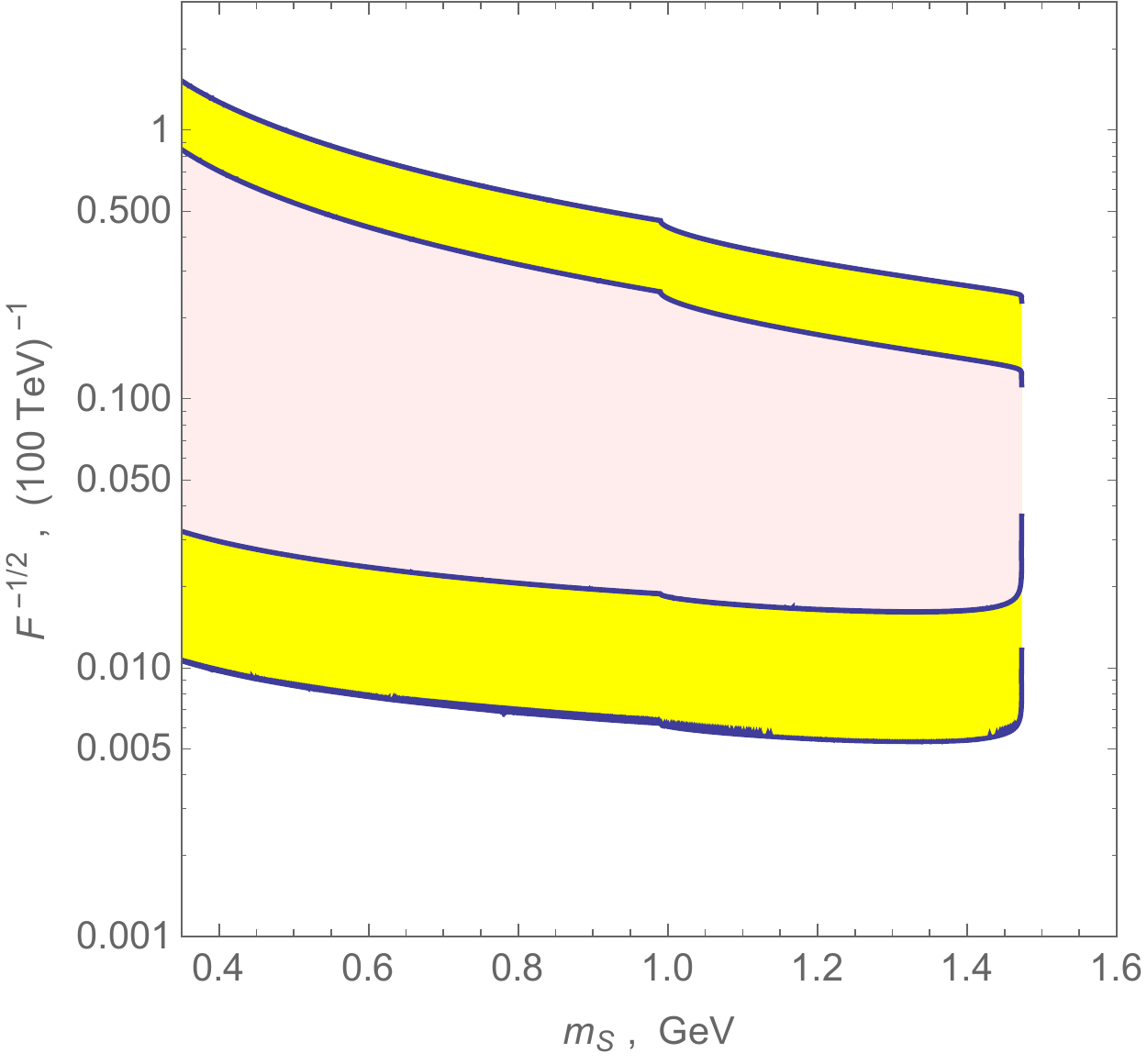}
\caption{\label{ScalarFV} Yellow shaded regions are expected to be excluded
(95\% C.L.)  at the SHiP experiment, if the flavor-violating soft parameters 
$\tilde{m}_{D_{23}}^{LR~2}$ (top panel) and
  $\tilde{m}_{U_{12}}^{LR~2}$ (bottom panel) take their present upper
  limits for the benchmark point in Table\,\ref{MSSMpoint}. 
The region above the solid horizontal line on the top panel is excluded
due to the negative result in searches for the three-body decay
$B\to K^0_s\nu\bar\nu$ \cite{Lutz:2013ftz}. 
Light red shaded regions are excluded by our analysis of the 
 results of the CHARM experiment \cite{Bergsma:1983rt,Bergsma:1985is}. 
}
\end{figure}

To illustrate the sensitivity of the SHiP experiment to the flavor-violating sgoldstino couplings, we take the soft parameters
$\tilde{m}_{U_{12}}^{LR~2}$ and $\tilde{m}_{D_{23}}^{LR~2}$ to be
equal to their upper bounds\,\eqref{phen-limits} 
and other MSSM parameters as in Table\,\ref{MSSMpoint}. Then,
treating the flavor-violating meson decays as the main sources of
sgoldstinos and repeating the analysis of Sec.\,\ref{subsec:D}, we
obtain the expected exclusion plots for the SHiP experiment (see 
Fig.\,\ref{ScalarFV}). 
Similar to Fig.\,\ref{Scalar}, the lower margins 
refer to the limit of tiny couplings of the sgoldstino 
to the SM fields. The numbers of sgoldstino decays here scale with the
relevant model parameters as 
$N_{\text{signal}}\propto
\tilde{m}_{D_{23}(U_{12})}^{LR~4}M_3^2/F^4$. 

Note that sgoldstinos can be directly searched for in the appropriate
decay modes of the heavy mesons (for examples, see 
Refs.\,\cite{Gorbunov:2000th,Gorbunov:2000cz,Demidov:2006pt,Demidov:2011rd}). 
In particular, the process $B\to K_s+S(P)$  with sgoldstinos
subsequently decaying outside the detector can mimic the process 
 $B\to h^{(*)}+\text{missing}$, whose branching is presently 
constrained as 
\cite{Lutz:2013ftz}  
\begin{equation}
\label{BK-br}
\text{Br}(B\to K^0_S \nu \bar\nu)<9.7\times 10^{-5}\,.
\end{equation}
At the chosen value of $\tilde{m}_{D_{23}}^{LR~2}$ the
sgoldstino's contribution (if any) must be smaller than this limit. 
A similar requirement exists for the twin process with charged
mesons. Thus, the region above  
the 
horizontal line in Fig.\,\ref{ScalarFV} (top panel) is excluded by the
constraint \eqref{BK-br}. This upper limit on $F^{-1/2}$ scales as
$\propto \text{Br}^{1/4}/\tilde{m}_{D_{23}}^{LR}$.   

Comparing Fig.\,\ref{ScalarFV} with Fig.\,\ref{Scalar} one concludes
that SHiP exhibits higher sensitivity to the supersymmetric models
with flavor violation.  We
extend the performed analysis to the CHARM experiment, which operated
on the same proton beam at CERN, but collected much lower statistics
and was placed at a  much larger distance from the target as compared to
the SHiP. Nevertheless, we find that in the case of flavor-violating
sgoldstino coupling, we can exclude a part of the model parameter
space (see Fig.\,\ref{ScalarFV}) given the negative results of
searches at CHARM~\cite{Bergsma:1983rt,Bergsma:1985is}.

\section{Pseudoscalar sgoldstino}
\label{sec:3}

If parity is (strongly) violated in the sfermion sector of the MSSM, sgoldstino
couplings to the SM fermions violate it, too. Then, scalar and
pseudoscalar sgoldstinos are very similar as regards the 
SHiP phenomenology. However, if sgoldstino couplings conserve parity 
(that takes place, e.g., in left-right
extensions of the MSSM), the phenomenology of pseudoscalar and scalar
sgoldstins is different in some aspects 
(see Refs.\,\cite{Gorbunov:2000th,Gorbunov:2000cz} for details). 
In this section we 
investigate SHiP sensitivity to the pseudoscalar sgoldstino couplings.

\subsection{Light pseudoscalar production}
\label{subsec:Abis} 

The pseudoscalar sgoldstino $P$ can be directly produced via the gluon
fusion with the same cross section as the scalar sgoldstino; hence,
Fig.\,\ref{Sigma} is valid for both cases.  However, its production
through the meson decays is somewhat different from that of the scalar
sgoldstino because of the absence of mixing with the light 
MSSM Higgs.\footnote{There is a mixing with $CP$-odd Higgs $A^0$, which
  is negligibly small for our choice of the benchmark point 
in Table\,\ref{MSSMpoint}.}

The light pseudoscalar can be produced in $B$-meson decays. The
corresponding one-loop diagram is very similar to that in 
the case of the scalar sgoldstino discussed in Sec.\,\ref{subsec:B}, 
but only the sgoldstino-top-top pseudoscalar coupling
\eqref{eef} contributes. Given the pseudoscalar nature of $P$, for the
two-body decay it is accompanied by the vector kaon $K^*$. The decay
rate can be obtained by properly replacing  the coupling
constants in the result presented in Ref.\,\cite{Freytsis:2009ct} 
for the case of decay into the axion, where we assume charged the Higgs boson mass $m_H$ to be approximately 1 TeV, which follows from the value of $m_A$ given in Table~\ref{MSSMpoint},
\begin{multline}
\Gamma(B\to{}K^*P)=\frac{G_F^2m_t^2}{2^{13}\pi^3}\frac{{m^{RL}_{U_{33}}}^4}
{2\,F^2}\cot^2\beta(\hat X_1+\cot^2\beta\hat X_2)^2\times\\
\times\frac{A_0^2\lambda^3_{B\to{}K^*}}{m_B^3},
\end{multline}
where 
\begin{equation}
\lambda_{B\to{}K^*}=\sqrt{(m_B^2-m_P^2-m_{K^*}^2)^2-4m_P^2m_{K^*}^2},
\end{equation}
and  
\begin{multline}
\hat X_1=2+\frac{m_H^2}{m_H^2-m_t^2}-\frac{3m_W^2}{m_t^2-m_W^2}+\\
+\frac{3m_W^4(m_H^2+m^2_W-2m_t^2)}{(m_H^2-m_W^2)(m_t^2-m^2_W)^2}\,\ln\frac{m_t^2}{m_W^2}+\\
+\frac{m_H^2}{m_H^2-m_t^2}\left(\frac{m_H^2}{m_H^2-m_t^2}-\frac{6m_W^2}{m_H^2-m_W^2}\right)\ln\frac{m_t^2}{m_H^2}\,,
\end{multline}

\begin{multline}
\hat X_2=-\frac{2m_t^2}{m_H^2-m_t^2}\left(1+\frac{m_H^2}{m_H^2-m_t^2}\ln\frac{m_t^2}{m_W^2}\right)\,,
\end{multline}
and the dimensionless form factor $A_0$ 
can be found in Ref.\,\cite{Ball:2004rg}. Though the formula for the
decay rate somewhat differs from that in the case of the scalar, 
they give very close numerical results, so the production rates are
almost the same, Figure\,\ref{SigmaBmeson} refers to both
cases. 

\subsection{Decay}
\label{subsec:Bbis}
 
In contrast to the scalar $S$, the pseudoscalar sgoldstino $P$ does not decay
into a meson pair. However, it mixes with pseudoscalar mesons
$\pi$ and $\eta$, as explained in Ref.\,\cite{Gorbunov:2000th}. Since
mesons exhibit four-meson coupling, the pseudoscalar sgoldstino can decay
into three mesons through the virtual meson state,   
$P\to{\pi^0}^*/\eta^*\to\text{3 mesons}$.

Inherent in the chiral perturbation theory the four-meson interaction 
reads~\cite{Pich:1998xt}
\begin{equation}
{\cal L}_{eff}=\frac{1}{12f_\pi^2}\Tr(\Phi\overleftrightarrow{\partial}\Phi\Phi\overleftrightarrow{\partial}\Phi),
\end{equation}
where
\begin{equation}
\Phi=\left(
\begin{matrix}
\frac{1}{\sqrt{2}}\pi^0+\frac{1}{\sqrt{6}}\eta & \pi^+ & K^+ \\
\pi^- & -\frac{1}{\sqrt{2}}\pi^0 +\frac{1}{\sqrt{6}}\eta & K^0 \\
K^- & \bar{K}^0 & -\frac{2}{\sqrt{6}}\eta 
\end{matrix}
\right).
\end{equation}
It induces 4-meson operators ${\cal O}_4^{\pi^0(\eta)}$ and ${\cal
  O}_4^{\eta}$ responsible for the interesting 
transitions ${\pi^0}^*/\eta^*\to\text{3 mesons}$. 

Matrix elements of the initial off-shell $\pi^0$ meson (with squared
4-momentum $m^2_{{\pi^0}^*}$) and 
three on-shell mesons  in the final state are as follows: 
\begin{align}\label{m1}
&\langle{\pi^0}^*|{\cal O}_4^{\pi^0}|\pi^0\pi^0\pi^0\rangle=\frac{1}{6f^2}(5m^2_{{\pi^0}^*}-3m^2_{\pi^0})\,,\\
\label{m2}
&\langle{\pi^0}^*|{\cal O}_4^{\pi^0}|\pi^0\eta\eta\rangle=\frac{2}{3f^2}(2m^2_{{\pi^0}^*}+m^2_{\pi^0}-2m^2_{\eta})\,,\\
\label{m3}
&\langle{\pi^0}^*|{\cal
  O}_4^{\pi^0}|\pi^0KK\rangle=\frac{2}{3f^2}(2m^2_{{\pi^0}^*}+m^2_{\pi^0}-2m^2_{K})\,,\\
\label{m4}
&\langle{\pi^0}^*|{\cal O}_4^{\pi^0}|\pi^0\pi^{\!+}\pi^{\!-}\rangle
=\frac{2}{3f^2}(2m^2_{{\pi^0}^*}+m^2_{\pi^0}-2m^2_{\pi^+})\,.
\end{align}
One obtains similar expressions for the initial virtual
$\eta$ meson. 

Finally we find for the pseudoscalar decay rate into 3 mesons:
\begin{multline}
\Gamma(P\to{\pi^0}^*/\eta^*\to\mbox{3 mesons})=\\
=\frac{f_\pi^2\pi^2m^4_{\pi}\epsilon^2}{4(m_P^2-m_{\pi})^2}\frac{M^2_3}{F^2}\Gamma(\pi^*\to\mbox{3 mesons})+\\
+\frac{f_\pi^2\pi^2m_{\eta}^4}{4(m_P^2-m_{\eta}^2)^2}\frac{M^2_3}{F^2}\Gamma(\eta^*\to\mbox{3 mesons}),
\end{multline}
where 3-body decay widths $\Gamma(\pi^*(\eta^*)\to\mbox{3 mesons})$
are calculated using the matrix elements ~\eqref{m1}~--~\eqref{m4} (and
similarly for the $\eta$ meson) assuming
that squared 4-momenta of the off-shell $\pi^*$ and $\eta^*$ equal
$m_P^2$; we also make use of $\epsilon=(m_u-m_d)/(m_u+m_d)$. 

Pseudoscalar sgoldstino decay rates into photons and leptons have 
the forms \eqref{Sgammagamma} and \eqref{Sll}, respectively, with the obvious
replacement $m_S\to m_P$.  

\begin{figure}[!htb]
    \centering
\includegraphics[width=0.45\textwidth]{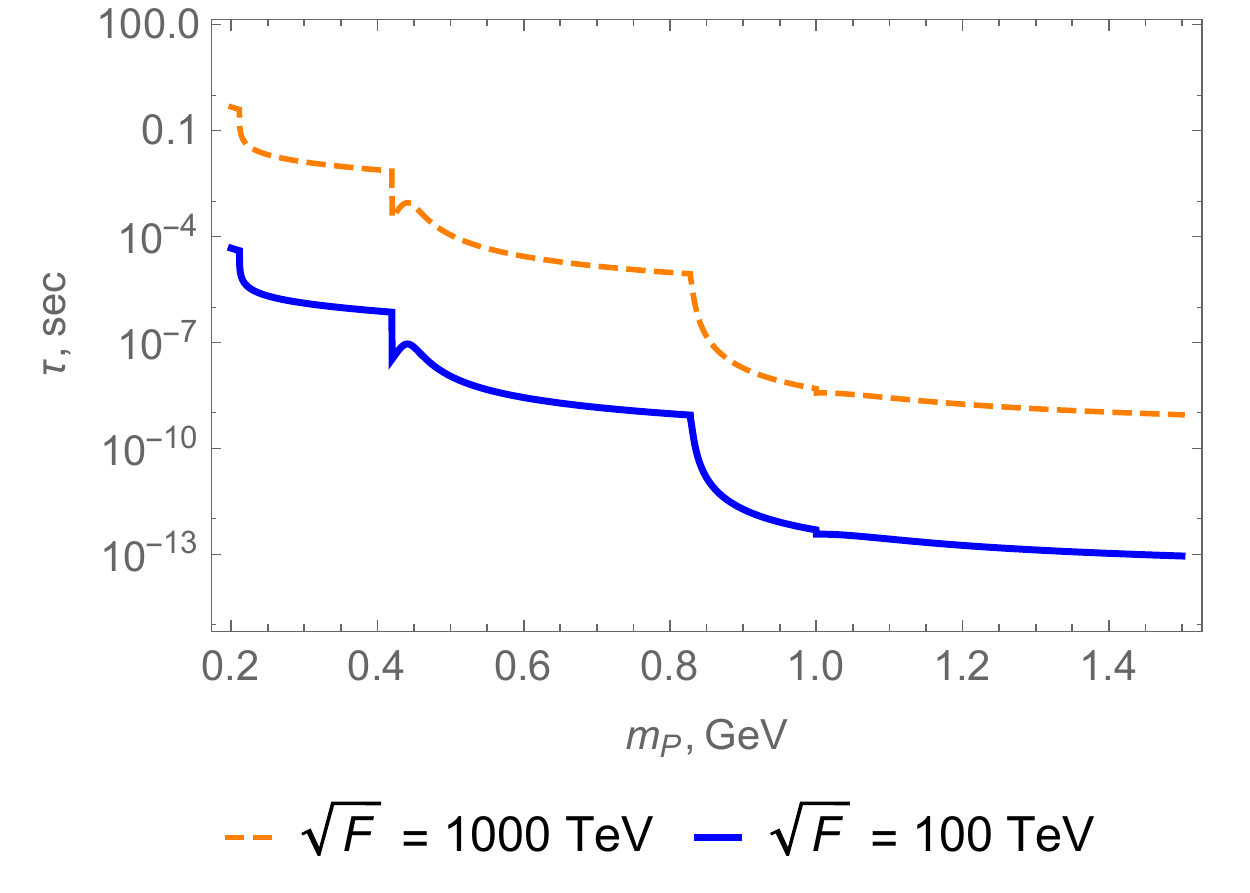}
    \caption{Pseudoscalar sgoldstino lifetime for
      $\sqrt{F}=100,\,1000$\,TeV (lines from bottom to top). 
    \label{PLifetime}
}
\end{figure}

\begin{figure}[!htb]
    \centering
\includegraphics[width=0.45\textwidth]{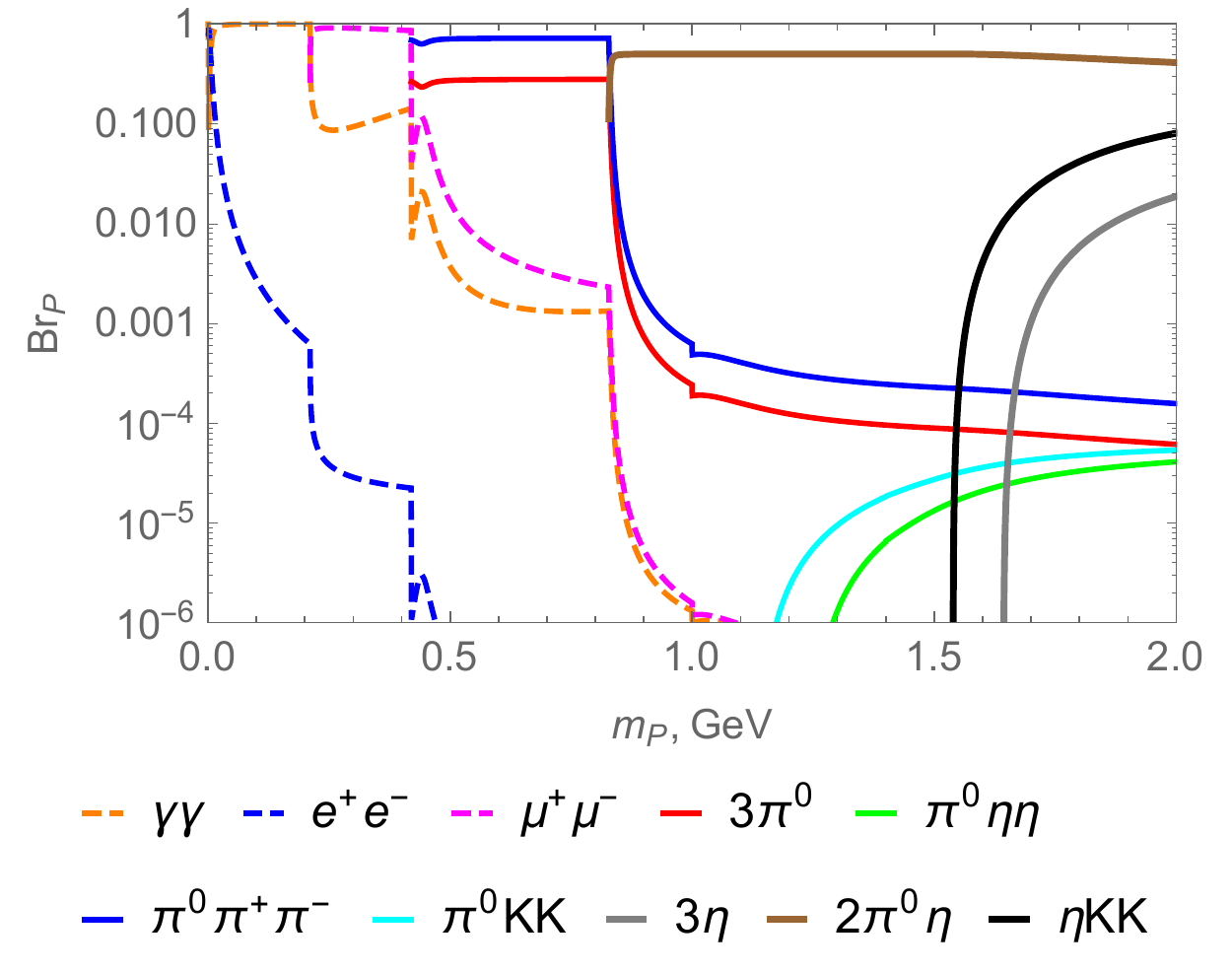}
    \caption{Branchings ratios of the pseudoscalar sgoldstino.
    \label{PBranching}}
\end{figure}

The pseudoscalar sgoldstino lifetime and its relevant decay branching
ratios are presented in Figs.~\ref{PLifetime} 
and \ref{PBranching}, 
correspondingly. Hadronic channels, mainly $3\pi$, $\eta KK$, $\eta\pi\pi$ and
$3\eta$, dominate sgoldstino decay. Given its geometry, the SHiP
experiment is sensitive to the supersymmetry breaking scales of about
$\sqrt{F}\sim 100$\,TeV and above.

\begin{figure}[!htb]
    \centering
\includegraphics[width=0.45\textwidth]{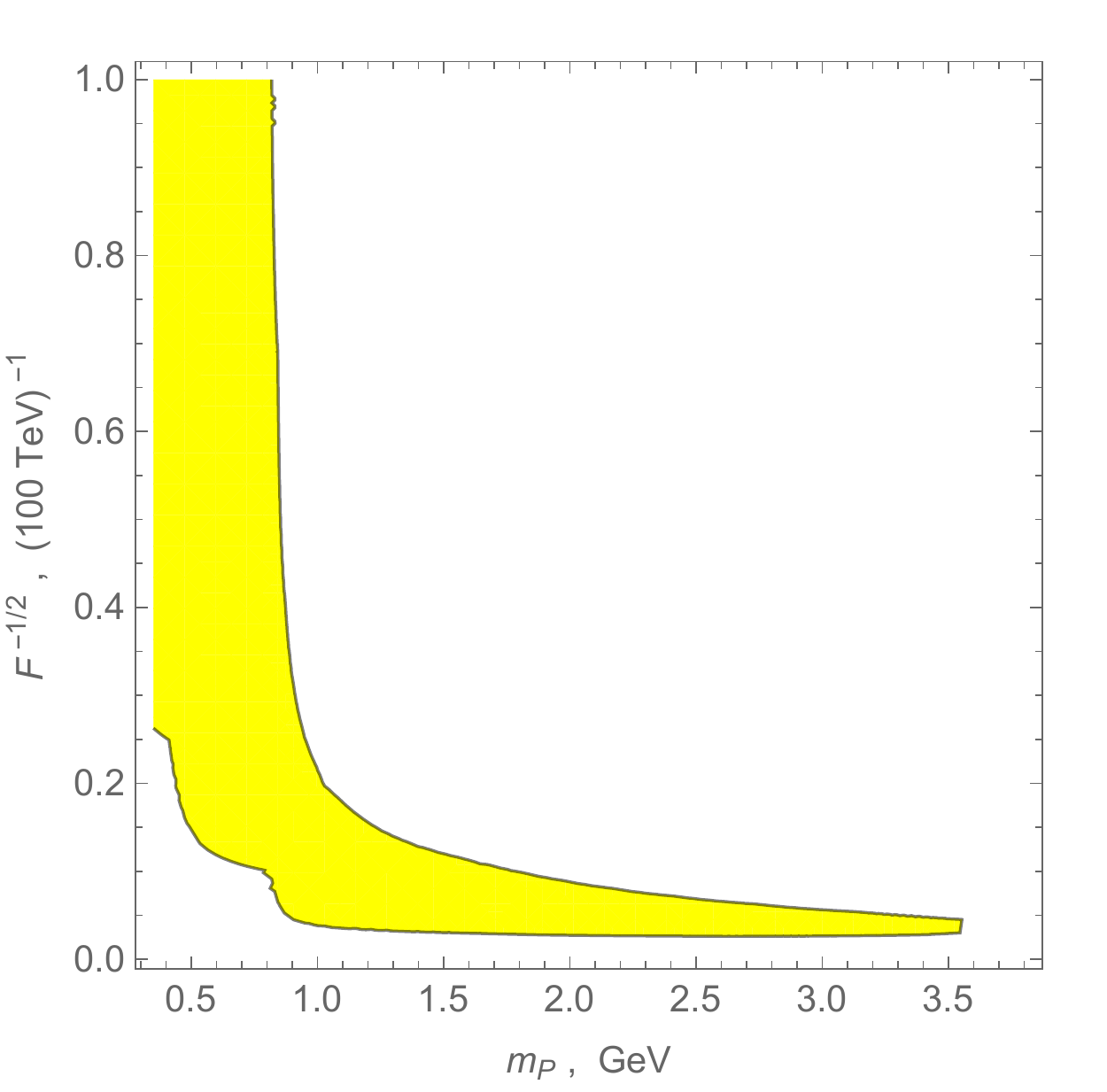}
    \caption{The shaded region can be explored at the SHiP experiment.}
    \label{Pseudoscalar}
\end{figure}

Performing for the pseudoscalar case the same procedure as that adopted in
Sec.\,\ref{subsec:D} for the scalar sgoldstino, we estimate 
the SHiP sensitivity to the pseudoscalar sgoldstino interaction. 
In Fig.~\ref{Pseudoscalar} 
we present the plot displaying the region in the parameter space
$(F^{-1/2},m_S)$ expected to be explored with the SHiP experiment. 
One observes that in the case of the pseudoscalar sgoldstino the 
SHiP is sensitive to models with a lower SUSY breaking scale 
$\sqrt{F}$ as compared to the scalar sgoldstino (cf. Figs.\ref{Scalar} and
\ref{Pseudoscalar}).

\subsection{Flavor violation}
\label{subsec:Cbis}
\begin{figure}[!htb]
\centering
\includegraphics[width=0.45\textwidth]{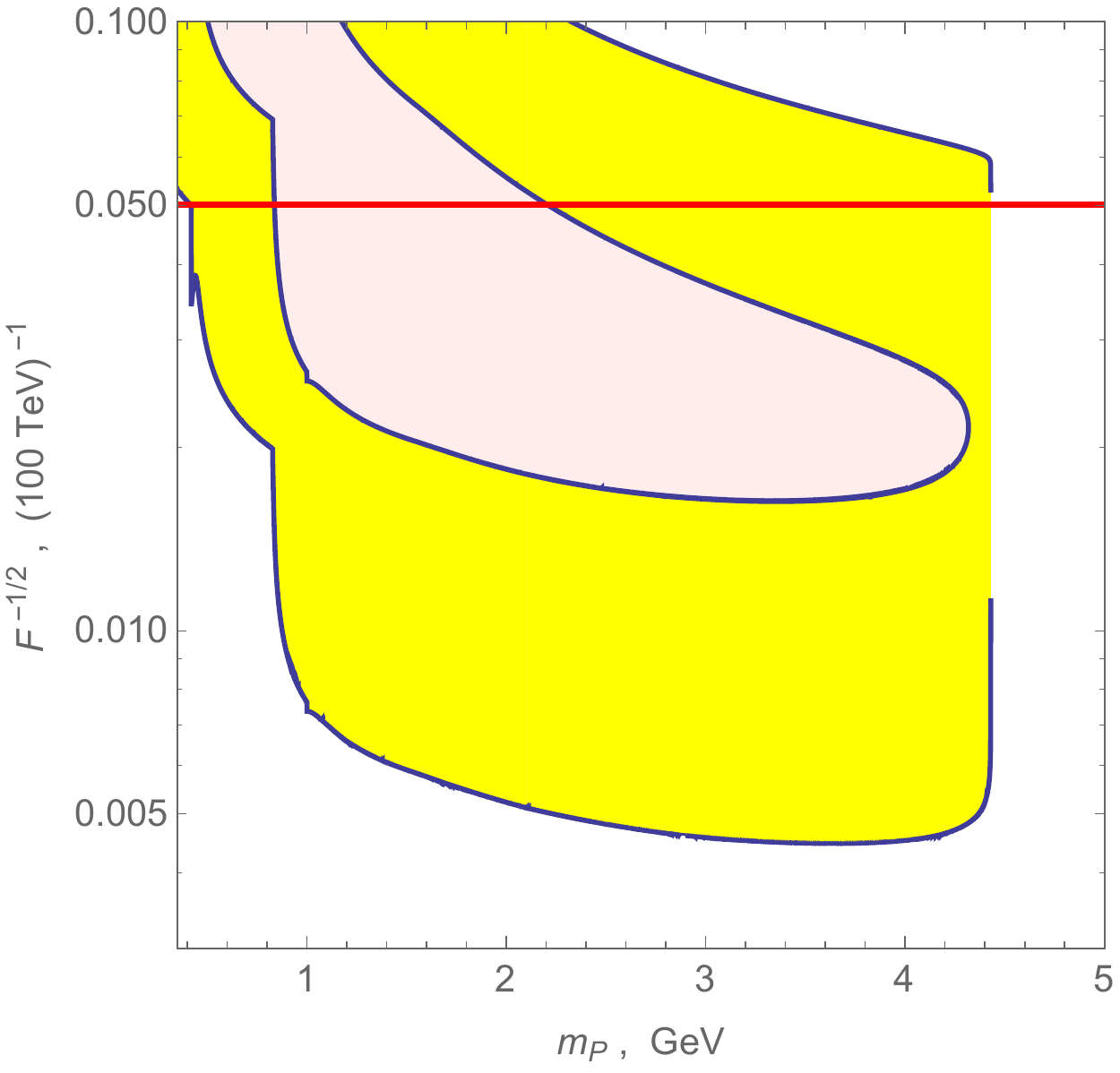}
\includegraphics[width=0.45\textwidth]{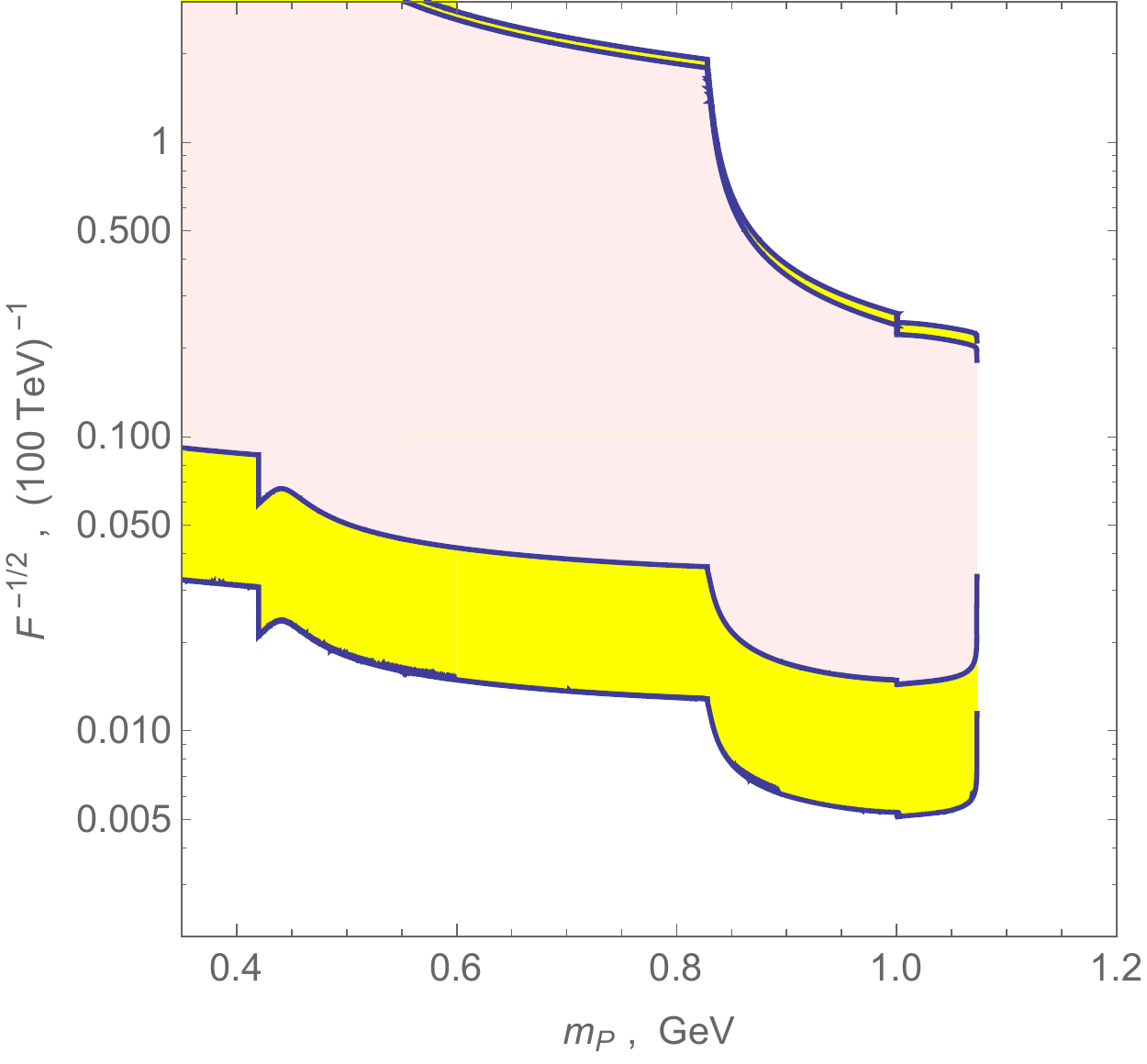}
\caption{\label{PseudoscalarFV} Yellow shaded regions can be explored
  with the SHiP experiment in the case of nonzero  soft 
parameters $\tilde{m}_{U_{12}}^{LR~2}$
  (lower panel) and $\tilde{m}_{D_{23}}^{LR~2}$ ( upper panel). 
The region above the solid horizontal line on the lower panel is excluded
due to a negative result in searches for the three-body decay
$B\to K^0_s\nu\bar\nu$ \cite{Lutz:2013ftz}. 
Light red shaded regions are excluded by the analysis of the 
 results of CHARM experiment \cite{Bergsma:1983rt,Bergsma:1985is}.
}
\end{figure}

The study of the flavor-violating pseudoscalar sgoldstino coupling is
very similar to that of the scalar sgoldstino. These couplings induce
two-body heavy meson decays into the pseudoscalar sgoldstino and a
light vector meson (it replaces the pseudoscalar meson in the case of the scalar
sgoldstino in the final state). For numerical estimates we adopt 
the same patterns of the flavor-violating couplings as used 
in the case of the scalar sgoldstino.  
The final estimates of the SHiP
sensitivity to the pseudoscalar sgoldstino are presented in
Fig.\,\ref{PseudoscalarFV}.

\section{Conclusions}
\label{sec:4}

We have estimated sensitivity of the SHiP experiment to supersymmetric
extensions of the SM where sgoldstinos are light. The proposed experiment
will be able to probe the supersymmetry breaking scale $\sqrt{F}$ up to
$10^3$\,TeV for the model without flavor violation and up to $10^5$\,TeV
for the model with flavor-violating parameters as large as the
corresponding present experimental upper bounds.  We have also
compared the regions of the sgoldstino parameter space to be probed at the SHiP
experiment with the regions that we have excluded from the analysis of the
results of the CHARM
experiment~\cite{Bergsma:1983rt,Bergsma:1985is}. The regions
are outlined in Figs.~\ref{ScalarFV} and \ref{PseudoscalarFV}: as
one can see, the SHiP experiment will significantly extend our ability
in testing supersymmetric models with light sgoldstinos. 

In this paper
we concentrate mostly on the models with sgoldstino masses in 
the range $0.4$--$4$\,GeV,
where sgoldstino can be kinematically produced in decays of charm and beauty
mesons. Lighter sgoldstinos can, in addition, be produced at the SHiP by
strange meson decays. This possibility deserves a special study beyond
the scope of this paper. Sgoldstinos can also be produced in decays of
secondary mesons from the hadronic cascade developing in the
target. These mesons are numerous, but less energetic. The proper
account of this contribution requires numerical simulations of the 
hadronic cascade and a more accurate treatment of the SHiP geometry.

\paragraph*{Acknowledgments}
We thank I.~Timiryasov and S.~Demidov for valuable discussions. The
work was supported by the RSF Grant No. 14-22-00161.



\begin{thebibliography}{99}

\bibitem{Haber:1984rc}
  H.~E.~Haber and G.~L.~Kane,
  ``The Search for Supersymmetry: Probing physics beyond the standard model,''
  Phys.\ Rep.\  {\bf 117}, 75 (1985).

\bibitem{Martin:1997ns} 
  S.~P.~Martin,
  ``A Supersymmetry primer,
  In *Kane, G.L. (ed.): Perspectives on supersymmetry II* 1-153
  [hep-ph/9709356].



\bibitem{Giudice:1998bp}
  G.~F.~Giudice and R.~Rattazzi,
  ``Theories with gauge mediated supersymmetry breaking,''
  Phys.\ Rep.\  {\bf 322} (1999) 419
  [hep-ph/9801271],

\bibitem{Dubovsky:1999xc}
  S.~L.~Dubovsky, D.~S.~Gorbunov and S.~V.~Troitsky,
  ``Gauge mechanism of mediation of supersymmetry breaking,''
  Phys.\ Usp.\  {\bf 42} (1999) 623
   [Usp.\ Fiz.\ Nauk {\bf 169} (1999) 705]
  [hep-ph/9905466].

\bibitem{Anelli:2015pba} 
  M.~Anelli {\it et al.} [SHiP Collaboration],
  ``A facility to Search for Hidden Particles (SHiP) at the CERN SPS,''
  arXiv:1504.04956 [physics.ins-det].



\bibitem{Gninenko:2013tk} 
  S.~N.~Gninenko, D.~S.~Gorbunov and M.~E.~Shaposhnikov,
  ``Search for GeV-scale sterile neutrinos responsible for active neutrino oscillations and baryon asymmetry of the Universe,''
  Adv.\ High Energy Phys.\  {\bf 2012}, 718259 (2012)
  [arXiv:1301.5516 [hep-ph]].


\bibitem{Bonivento:2013jag} 
  W.~Bonivento {\it et al.},
  ``Proposal to Search for Heavy Neutral Leptons at the SPS,''
  arXiv:1310.1762 [hep-ex].



\bibitem{Alekhin:2015byh} 
  S.~Alekhin {\it et al.},
  ``A facility to Search for Hidden Particles at the CERN SPS: the SHiP physics case,''
  arXiv:1504.04855 [hep-ph].




\bibitem{Cremmer:1978iv}
  E.~Cremmer, B.~Julia, J.~Scherk, P.~van Nieuwenhuizen, S.~Ferrara and L.~Girardello,
 ``Super-higgs effect in supergravity with general scalar interactions,''
  Phys.\ Lett.\ B {\bf 79} (1978) 231.




\bibitem{Gorbunov:2000th} 
  D.~S.~Gorbunov,
  ``Light sgoldstino: Precision measurements versus collider searches,''
  Nucl.\ Phys.\ B {\bf 602}, 213 (2001)
  [hep-ph/0007325].

\bibitem{Gorbunov:2001pd} 
  D.~S.~Gorbunov and A.~V.~Semenov,
  ``CompHEP package with light gravitino and sgoldstinos, 
  hep-ph/0111291.


\bibitem{Perazzi:2000id} 
  E.~Perazzi, G.~Ridolfi and F.~Zwirner,
  ``Signatures of massive sgoldstinos at e+ e- colliders,''
  Nucl.\ Phys.\ B {\bf 574}, 3 (2000)
  [hep-ph/0001025].

\bibitem{Perazzi:2000ty} 
  E.~Perazzi, G.~Ridolfi and F.~Zwirner,
  ``Signatures of massive sgoldstinos at hadron colliders,''
  Nucl.\ Phys.\ B {\bf 590}, 287 (2000)
  [hep-ph/0005076].


\bibitem{Demidov:2011rd} 
  S.~V.~Demidov and D.~S.~Gorbunov,
  ``Flavor violating processes with sgoldstino pair production,''
  Phys.\ Rev.\ D {\bf 85}, 077701 (2012)
  [arXiv:1112.5230 [hep-ph]].



\bibitem{Dudas:2012fa} 
  E.~Dudas, C.~Petersson and P.~Tziveloglou,
 ``Low Scale Supersymmetry Breaking and its LHC Signatures,''
  Nucl.\ Phys.\ B {\bf 870}, 353 (2013)
  [arXiv:1211.5609 [hep-ph]].


\bibitem{Bellazzini:2012mh} 
  B.~Bellazzini, C.~Petersson and R.~Torre,
  ``Photophilic Higgs from sgoldstino mixing,''
  Phys.\ Rev.\ D {\bf 86}, 033016 (2012)
  [arXiv:1207.0803 [hep-ph]].

\bibitem{Astapov:2014mea} 
  K.~O.~Astapov and S.~V.~Demidov,
  ``Sgoldstino-Higgs mixing in models with low-scale supersymmetry breaking,''
  JHEP {\bf 1501}, 136 (2015)
  [arXiv:1411.6222 [hep-ph]].

\bibitem{Ellis:1979jy}
J.~R. Ellis, M.~K. Gaillard, D.~V. Nanopoulos, and C.~T. Sachrajda, 
 Phys.\ Lett.\ B {\bf 83}, 339 (1979).


\bibitem{Bezrukov:2009yw}
  F.~Bezrukov and D.~Gorbunov,
  ``Light inflaton Hunter's Guide,''
  JHEP {\bf 1005} (2010) 010
  [arXiv:0912.0390 [hep-ph]].



\bibitem{Gorbunov:2015mba} 
  D.~Gorbunov and I.~Timiryasov,
  ``Decaying light particles in the SHiP experiment. II. Signal rate estimates for light neutralinos,''
  Phys.\ Rev.\ D {\bf 92}, no. 7, 075015 (2015)
  [arXiv:1508.01780 [hep-ph]].


\bibitem{Ciuchini:2007cw} 
  M.~Ciuchini, E.~Franco, D.~Guadagnoli, V.~Lubicz, M.~Pierini, V.~Porretti and L.~Silvestrini,
  ``$D$ - $\bar{D}$ mixing and new physics: General considerations and constraints on the MSSM,''
  Phys.\ Lett.\ B {\bf 655}, 162 (2007)
  [hep-ph/0703204].

\bibitem{Arana-Catania:2013pia} 
  M.~Arana-Catania,
  ``The flavour of supersymmetry: Phenomenological implications of sfermion mixing,''
  arXiv:1312.4888 [hep-ph].

\bibitem{Palmer:2013yia} 
  T.~Palmer and J.~O.~Eeg,
  ``Form factors for semileptonic D decays,''
  Phys.\ Rev.\ D {\bf 89}, no. 3, 034013 (2014)
  [arXiv:1306.0365 [hep-ph]].


\bibitem{Lutz:2013ftz} 
  O.~Lutz {\it et al.} [Belle Collaboration],
  ``Search for $B \to h^{(*)} \nu \bar{\nu}$ with the full Belle $\Upsilon(4S)$ data sample,''
  Phys.\ Rev.\ D {\bf 87}, no. 11, 111103 (2013)
  [arXiv:1303.3719 [hep-ex]].


\bibitem{Bergsma:1983rt} 
  F.~Bergsma {\it et al.} [CHARM Collaboration],
  ``A Search for Decays of Heavy Neutrinos,''
  Phys.\ Lett.\ B {\bf 128}, 361 (1983).

\bibitem{Bergsma:1985is} 
  F.~Bergsma {\it et al.} [CHARM Collaboration],
  ``A Search for Decays of Heavy Neutrinos in the Mass Range 0.5-{GeV} to 2.8-{GeV},''
  Phys.\ Lett.\ B {\bf 166}, 473 (1986).

\bibitem{Gorbunov:2000cz} 
  D.~S.~Gorbunov and V.~A.~Rubakov,
  ``Kaon physics with light sgoldstinos and parity conservation,''
  Phys.\ Rev.\ D {\bf 64}, 054008 (2001)
  [hep-ph/0012033].

\bibitem{Demidov:2006pt} 
  S.~V.~Demidov and D.~S.~Gorbunov,
  ``More about sgoldstino interpretation of HyperCP events,''
  JETP Lett.\  {\bf 84}, 479 (2007)
  [hep-ph/0610066].


\bibitem{Freytsis:2009ct} 
  M.~Freytsis, Z.~Ligeti and J.~Thaler,
  ``Constraining the Axion Portal with B ---> K l+ l-,''
  Phys.\ Rev.\ D {\bf 81}, 034001 (2010)
  [arXiv:0911.5355 [hep-ph]].

\bibitem{Ball:2004rg} 
  P.~Ball and R.~Zwicky,
  ``B(D,S) ---> rho, omega, K*, phi decay form-factors from light-cone sum rules revisited,''
  Phys.\ Rev.\ D {\bf 71}, 014029 (2005)
  [hep-ph/0412079].

\bibitem{Pich:1998xt} 
  A.~Pich,
  ``Effective field theory: Course,''
  hep-ph/9806303.





\end{thebibliography}
\end{document}